\documentclass[aps,prl,superscriptaddress,showpacs,twocolumn]{revtex4}
\usepackage{graphicx}
\usepackage{color}
\begin{document}

\affiliation{Institute of Physics, P.O.B.304, HR-10 000, Zagreb, Croatia}

\title{Three Dimensional Spin Web: A New Magnetic Lattice in Cu$_{3}$TeO$_{6}$}

\author{M.Herak}
  \email{mirta@ifs.hr}
\affiliation{Institute of Physics, P.O.B.304, HR-10 000, Zagreb,
Croatia}
\author{H. Berger}
\affiliation{Institute de Physique de la Mati\`{e}re Complexe, EPFL, CH-1015 Lausanne, Switzerland}

\author{M.Prester}
  \email{prester@ifs.hr}
\affiliation{Institute of Physics, P.O.B.304, HR-10 000, Zagreb, Croatia}

\author{M. Miljak}
\affiliation{Institute of Physics, P.O.B.304, HR-10 000, Zagreb, Croatia}

\author{I. \v{Z}ivkovi\'{c}}
\affiliation{Institute of Physics, P.O.B.304, HR-10 000, Zagreb, Croatia}

\author{O. Milat}
\affiliation{Institute of Physics, P.O.B.304, HR-10 000, Zagreb, Croatia}

\author{D. Drobac}
\affiliation{Institute of Physics, P.O.B.304, HR-10 000, Zagreb, Croatia}

\author{S. Popovi\'{c}}
\affiliation{Faculty of Science, Physics Department, Bijeni\v{c}ka c. 32, HR-10 000 Zagreb, Croatia}

\author{O. Zaharko}
	\email{oksana.zaharko@psi.ch}
\affiliation{Laboratory for Neutron Scattering, ETHZ and PSI, CH-5232 Villigen, Switzerland}

\date{\today}
\begin{abstract}
We report on magnetic properties of cubic compound Cu$_3$TeO$_6$ studied by $ac$ and $dc$ susceptibility and neutron powder diffraction for the first time.
 A novel magnetic lattice, three dimensional spin web, composed of almost planar regular hexagons of Cu$^{2+}$ $1/2$ spins, defines the properties of Cu$_3$TeO$_6$. The behaviour of the magnetic susceptibility in the paramagnetic state at $\approx$ 170 K is suggestive for a competition between local anisotropies of Cu$^{2+}$ hexagons. The resulting frustration is weaker than the antiferromagnetic nearest-neighbor interaction which leads to a collinear (or slightly canted) spin arrangement (k=0,0,0) and formation of magnetic domains below $T_N= 61\:$K.\\
\end{abstract}



\maketitle

\begin{figure}[t!]
\begin{center}
\resizebox{0.9\columnwidth}{!}{
 \includegraphics{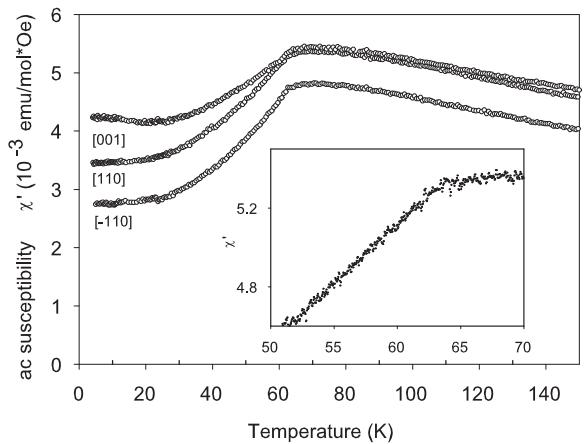}
}
\end{center} 
\caption{Main panel: ac susceptibility measurements on a single-crystalline Cu$_3$TeO$_6$ sample. Applied ac field ($2\:$Oe, $430\:$Hz) was directed along the specified crystallographic directions. The vertical downward shift for the [-110] sample orientation might originate from reasons specific to ac susceptibility technique. Inset: Magnetic transition on the expanded scale.}
\label{fig:1}
\end{figure}

\indent Magnetism relying on 3d$^9$ copper Cu$^{2+}$ ions reveals a striking diversity of magnetic structures. This diversity originates from a broad range of effective magnetic dimensionalities characterizing various 3d$^9$ magnetic systems \cite{review}. Depending, in turn, on the level of frustration and the importance of quantum fluctuations either singlet non-magnetic, disordered spin liquid or magnetically long-range ordered ground states set in at low temperatures. Understanding the magnetic structures underlying each of these ground states represents an issue of central interest for 3d$^9$-magnetism, as well for magnetism in general.\\
\indent In this article we present magnetic susceptibility and neutron diffraction studies of a copper tellurium oxide, Cu$_3$TeO$_6$. Apart from the crystal structure determination, published for the first time \cite{Hostachy:1968} back in 1968 and revised \cite{Falck:1978} in 1978, no other properties of Cu$_3$TeO$_6$ have, to the best of our knowledge, been ever reported in literature, in spite of tremendous general interest for magnetism of copper- and other transition metal-oxides. In this work we focus particularly on magnetic ordering which sets in at $61\:$K. Scrutinizing the crystallographic and neutron diffraction data of  Cu$_3$TeO$_6$, a novel type of magnetic lattice, hereafter referred to as {\em a three-dimensional spin web}, have been identified. In this lattice, almost planar neighboring Cu$^{2+}$ hexagons share one common corner and, by buckling and folding in space, form a complex 
3D network. Compared with a related 3D pyrochlore lattice \cite{Lee:2002} frustration in the Cu$_3$TeO$_6$ spin web has been found to be relatively weak, but it plays a key role in low temperature magnetism.\\
\indent Single crystals of Cu$_3$TeO$_6$ were obtained by HBr chemical transport method in sealed quartz tubes with temperature gradients of $600-550^\circ\:$C and $450-500^\circ\:$C respectively. The resulting prisms were $2\:$mm long and 1 to $2\:$mm wide. The room temperature X-ray powder diffraction patterns were collected on a Philips automatic diffractometer. They could be indexed in the cubic structure ($Ia\overline{3}$, $a=9.537(1)\:$\AA\space), reported \cite{Hostachy:1968, Falck:1978} previously.\\
\indent High resolution $ac$ susceptibility was measured in the temperature interval $4.2\:$K to $200\:$K by 
 a commercial CryoBIND system. The amplitude of $ac$ fields in the latter technique is usually small (at the order of $1\:$Oe, typically) representing an important advantage in studies of any kind of {\em spontaneous} magnetic ordering.  $dc$ susceptibility was measured by Faraday method in the temperature range from $2\:$K to $330\:$K in the field of $5\:$kG. Neutron powder diffraction patterns were collected in the temperature range $1.5\:$K- $70\:$K on the DMC diffractometer at SINQ, Switzerland, with neutron wavelength 2.568 \AA. \\
\indent The low-field $ac$ susceptibility is shown in Fig. \ref{fig:1}. A kink at about $61\:$K is a clear indication of 3d long-range magnetic ordering, presumably of an antiferromagnetic origin. Compared with classical antiferromagnets revealing N\'{e}el order one notes the absence of pronounced susceptibility anisotropy, otherwise characteristic for uniform N\'{e}el-ordered systems. This observation might be consistent with more complex (e.g., helicoidal) AF ordering. However, small susceptibility anisotropy can most naturally be interpreted by presence of AF domains of different orientations, rendering the effective bulk susceptibility more isotropic. The latter interpretation turns out to be consistent with all other observations presented below.
\\
\begin{figure}[b!]
\begin{center}
\resizebox{0.9\columnwidth}{!}{
 \includegraphics{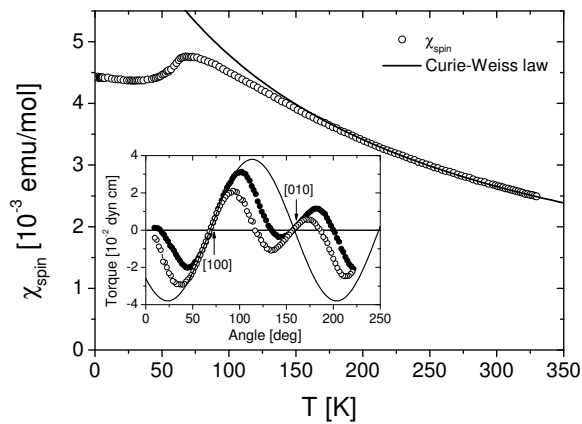}
}
\end{center} 
\caption{$dc$ magnetic susceptibility of Cu$_3$TeO$_6$ in the applied field of $5\:$kOe. Inset: Torque measurement in field of 1.1 kOe at 4.2 K on the same sample. Applied dc field was swept from $\approx$ 10$^{\circ}$ up to 220$^{\circ}$ (solid circles) and than back (empty circles). Positions of the crystal axes [100] and [010] are  marked. Solid line represents the $sin(2 \theta)$ curve. 
}
\label{fig:2}
\end{figure}
\indent Figure \ref{fig:2} shows the total  $dc$ magnetic susceptibility in a broad temperature range. The spin-only part of susceptibility $\chi_{spin}$ was calculated by subtracting the temperature independent part $\chi_{0}$. $\chi_{0}$ summs the diamagnetic susceptibility of all ions and the paramagnetic {Van Vleck} contribution of Cu$^{2+}$, thus $\chi_{0}=+1.5\cdot10^{-4}\:$emu/mol. At approximately $170\:$K a deviation from Curie-Weiss law sets in.  Inset to Figure \ref{fig:2} shows that $\chi_{spin}$ follows the Curie-Weiss law, $1/\chi_{spin} = (T-\Theta_{CW})/C$, in the temperature range $170-330\:$K. The fit gives the values $C=1.18\:$emuK/mol and $\Theta_{CW}\approx-145\:$K for the Curie and the Weiss constant, respectively. Large and negative $\Theta_{CW}$ from the Curie-Weiss law suggests that the copper spins are strongly antiferromagnetically coupled. The value of the g-factor calculated from the determined $C$ equals $g=2.05$. This value is smaller than the value of $<g> \approx 2.15$ characterizing a large number of investigated copper oxides. This suggests there is an approximately 5\% copper spin-deficiency in our sample. Alternatively, a g-factor deviation could rely on inappropriate use of Curie-Weiss law even at the highest measured temperatures (inset to Figure \ref{fig:2}). Below $\approx 170\:$K the susceptibility increases less rapidly than the original Curie-Weiss law, reveals a maximum at $69\:$K and then decreases rapidly below $61\:$K.\\

\begin{figure}[t!]
\begin{center}
\resizebox{0.9\columnwidth}{!}{
 \includegraphics{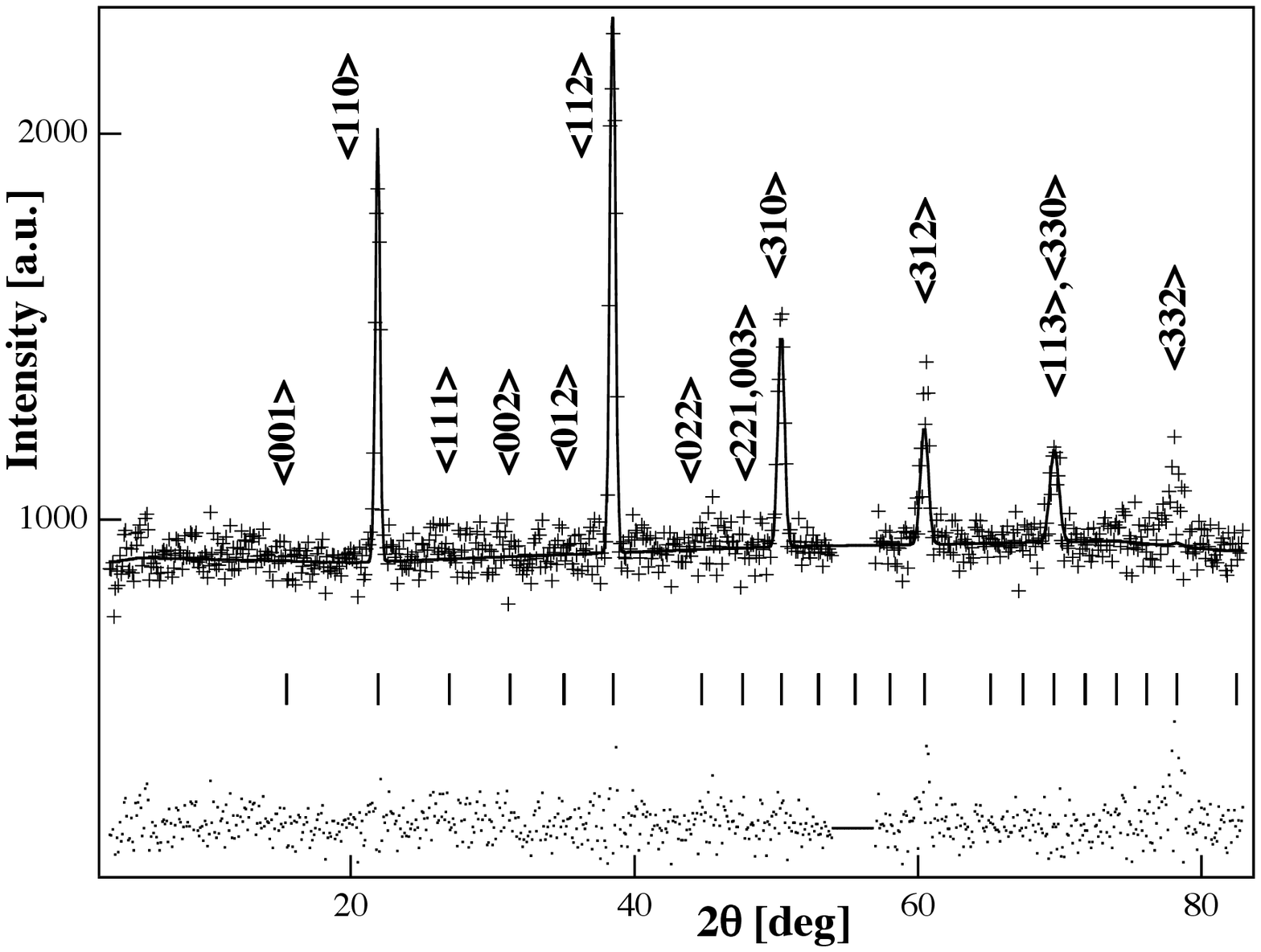}
}
\end{center} 
\caption{Observed 1.5 K-70 K magnetic difference pattern, calculated and difference patterns of Cu$_3$TeO$_6$ denoted by crosses, solid, and dotted lines, respectively. Note, the indices of reflections are cyclically permutable.}
\label{fig:4}
\end{figure}

\begin{figure}[t!]
\begin{center}
\resizebox{0.85\columnwidth}{!}{
 \includegraphics{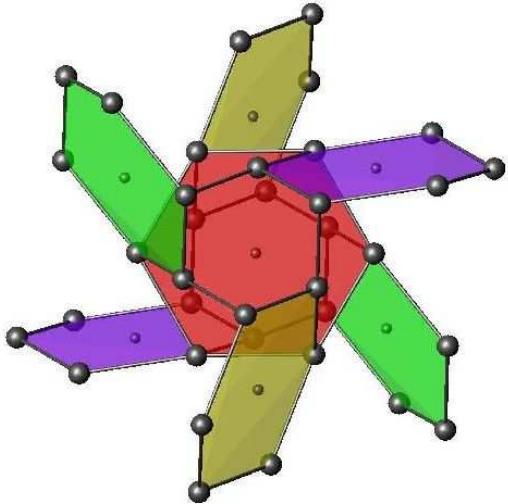}
}
\end{center} 
\caption{Growth of the spin web out the single (red) hexagon by three-dimensional hexagon buckling. Each copper ion (large black spheres) is coordinated by four nearest neighbors. There are four pairs of non-equivalent hexagons shown in different colors.}
\label{fig:5}
\end{figure}

\begin{figure}[t!]
\begin{center}
\resizebox{0.8\columnwidth}{!}{
 \includegraphics{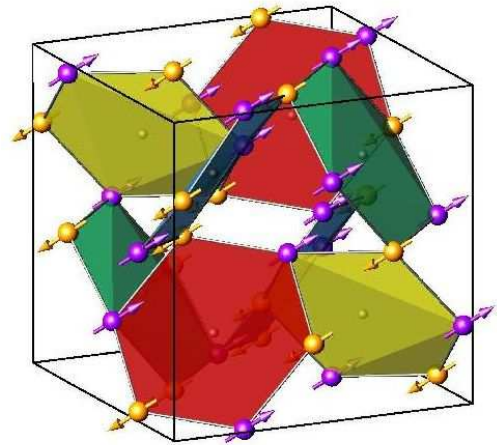}
}
\end{center} 
\caption{Three-dimensional spin web, as determined by a network of copper hexagons and collinear spin arrangement along one (out of four allowed by symmetry) body diagonal. Hexagons are formed around TeO$_6$ octahedra (only tellurium ions in the middle of the hexagons are shown). There are four pairs of non-equivalent hexagons in the unit cell, shown in different colors.}
\label{fig:6}
\end{figure}

\indent  One notes that magnetic ordering introduces susceptibility reduction being almost 5 times bigger in $ac$- compared to the $dc$- susceptibility studies. Taking into account that the respective measuring fields differ by 3 orders of magnitude one concludes that a pronounced and unusual magnetic non-linearity characterizes the ordered phase \cite{choi}. \\ 
\indent Torque magnetometry studies (Inset to Fig.\ref{fig:2}) reveal a strong deviation from those characterizing single domain antiferromagnet, i.e., sinusoidal angle dependence with $\pi$ periodicity and zeros in the direction of the magnetic axes. In Cu$_{3}$TeO$_{6}$ this type of behavior was observed only in low fields ($<0.5$ kOe). In higher fields, instead of a narrow and non-hysteretic spin flop, we observed a sequence of discontinuous jumps and a pronounced hysteresis. Closely related are the obervations of the relaxation effects on a long time scale ($\tau \approx 200$ seconds). As the same behavior was observed also for another orientation of the sample one concludes there are at least
four AF  domains which change their population at fields $>0.5$ kG. The details of these studies will be published separately \cite{Milj}.\\%
\begin{figure}[b!]
\begin{center}
\resizebox{0.9\columnwidth}{!}{
 \includegraphics{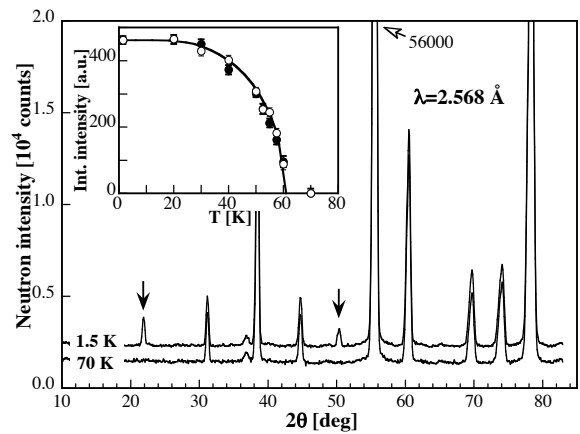}
}
\end{center} 
\caption{DMC neutron powder diffraction patterns of Cu$_3$TeO$_6$. Arrows points to magnetic reflections.  Inset shows the temperature evolution of the $\langle hk0\rangle$ (opened circles) and $\langle hkl\rangle$ (filled circles) lowest angle magnetic peaks intensity.}
\label{fig:3}
\end{figure}
\indent Below T$_N = 61$ K magnetic peaks appear in neutron powder diffraction, Figure \ref{fig:3}, at the positions of the crystallographic reciprocal lattice corresponding to the wave vector k=(0,0,0). The temperature variation of the integrated intensity of the magnetic peaks has classical behaviour; the intensities of the $<hkl>$ and $<hk0>$ contributions vary as the square of the S=1/2 Brillouin function.\\
\indent The systematic extinction rules observed in 1.5 K-70 K magnetic difference pattern (Figure \ref{fig:4}) reveal that the I-translation is not combined with time reversal ($hkl$: $h+k+l = 2n$), while the glide planes
are combined with it, if retained ($<hk0>$: $<k,l> = 2n$;  $<hkl>$ denotes cyclical permutation). Note that three-fold rotations are not compatible with time reversal. Our trials to find a model with the cubic magnetic configuration symmetry were unsuccessful. Only models with trigonal symmetry gave good agreement to the observed diffraction pattern. The retention of the unique three-fold axis in the magnetic structure has two consequences. Firstly, there must exist at least four rotation S-domains, each possessing its own $<1 \pm 1 \pm 1>$ three-fold axis and each having a pair of 180 deg domains. Their presence is fully supported by the torque magnetometry results. Secondly, as the cubic symmetry is lost, the glide planes could not be the elements of the magnetic group. The Cu$^{2+}$ ions related by the glide planes must, however, have the magnetic moments antiparallel to each other, otherwise magnetic intensity would be found at the $hkl$: $h + k + l = 2n$ positions.\\
\indent The magnetic moment direction is not conditioned by extinctions and must be determined from modeling. Due to the high symmetry of the crystal lattice and the wave vector this task is, however, not easy based on powder data
only \cite{Shirane:1958}. For a collinear antiferromagnetic model with spins aligned along the [111] direction and the 1.5 K moment value of 0.644(7) $\mu_B$/Cu$^{2+}$ a good fit (R$_M=16.9$\%) has been obtained. However, canted spin arrangements with magnetic moments tilted from the [111] direction, fit the data equally well (R$_M=13.2$\%). For collinear model spins are aligned along one of the space diagonals of the cubic unit cell. Since each of the four directions is equally probable, domains with four different spin directions must coexist in antiferromagnetically ordered state. In canted model the tilt of the spins from the [111] direction is small, of the order of 6 degrees. Here also four equally probable antiferromagnetic domains could exist. In canted model the angle between two 1$^{st} \;nn$ moments is 168.7 degrees (in collinear model it is 180 deg) and between two 2$^{nd} \;nn$ 11.3 degrees (in collinear model it is 0 deg). Further neutron diffraction experiments on a single domain crystal are needed to distinguish between the models.\\
\indent Now we elaborate the magnetic structure of Cu$_3$TeO$_6$. 
There are 24 copper ions per unit cell, each forming a distorted CuO$_6$ octahedron. Each Cu${}^{2+}$ ion has four nearest Cu neighbours $(nn)$ at 3.18\AA\space ($1st$ $nn$) and the next four $nn$ at 3.6\AA\space ($2nd$ $nn$). $1st$ $nn$ are connected by 2 superexchange paths through two oxygens making Cu-O-Cu angles of 92.4$^{\circ}$ and 106.2$^{\circ}$, while $2nd$ $nn$ are connected through one oxygen only, forming a Cu-O-Cu angle of 112.5$^{\circ}$. Assuming that the interaction between $2nd$ $nn$ is weaker compared to the $1st$ $nn$ one, it seems reasonable to restrict consideration of magnetic interactions just to the $1st$ $nn$ network. \\
\indent The latter restriction generates a surprisingly interesting outcome, Figures \ref{fig:5},\ref{fig:6}. The structural building block of the $1st$ $nn$ sublattice is characterized by almost planar hexagon arrangement of copper ions. As shown in Fig. \ref{fig:5} the hexagons are not isolated, but form a complex three-dimensional network, hereby named {\em a three-dimensional spin web}. Note that the web geometry is primarily determined just by copper ion topology: as shown in Figure \ref{fig:5} each Cu${}^{2+}$ is coordinated by four nearest neighbors and is shared between the two non-coplanar hexagons \cite{fus1}.\\
\indent The results of neutron diffraction studies are indeed consistent with a collinear AF arrangement of spins within hexagons (Figs.\ref{fig:6}). The spin web with hexagons as the building blocks strongly resembles the pyrochlore lattice of corner-shared tetrahedra \cite{Lee:2002}. The latter lattice characterizes AB$_{2}$O$_{4}$ spinels, the cubic systems as well. There are, however, important differences: While in the pyrochlore the disconnected hexagons are spanned by a skeleton of spin tetrahedra, in the spin web the hexagons share a common corner and are interconnected by a network of distorted CuO$_6$ octahedra. (The Cu-O distances in each spin web octahedron are: 1.949(2)\AA\space (2x), 2.031(2)\AA\space (2x) and 2.369(3)\AA\space (2x) while the O-Cu-O angles in the octahedron range from 72.6(1)$^{\circ}$ to 166.3$^{\circ}$, none of them being 90$^{\circ}$ nor 180$^{\circ}$ as in a regular octahedron). \\
\indent In the pyrochlore  (in particular ZnCr$_{2}$O$_{4}$ \cite{Lee:2002}) the pronounced frustration relies on the geometrically frustrated spin tetrahedra building blocks. In the spin web Cu$_3$TeO$_6$ the possible source of frustration is the local magnetic anisotropy of the hexagons, in addition to the neglected interaction with $2nd$ nearest neighbors.  Each spin shared between the two non-coplanar hexagons experiences frustration. Apparently, this frustration is only modest (the value of frustration parameter $f=\Theta_{CW}/T_{N}=2.4$). Alternatively, one can say that the spin web is closer to the over-constrained limit \cite{Ramirez:2001}, preferring magnetic order at higher temperature.\\
\indent Furthermore, a comparison of the two lattices seems to reveal a route how frustration can affect magnetic clustering in general. In the pyrochlore strong frustration promotes clustering of individual spins into hexagonal loops at low temperatures \cite{Lee:2002}. The loop directors - the unique direction along which the spins are aligned - are weakly interacting and slowly varying in space. In the spin web the directors are strongly interacting. This interaction, mediated by a common corner, overcomes the frustration and the collinear order fixes the directors along the common axis. \\
\indent In conclusion, a new magnetic lattice, the 3d spin web, characterizing magnetic structure of the Cu$_3$TeO$_6$  compound has been found. It's main building blocks, Cu hexagons, share common corners. The competition between local anisotropy of hexagons and AF nearest neighbors interaction leads to a modest frustration which is resolved below  $T_{N}=61\:$ K by formation of AF collinear spin arrangement.
The unusual magnetic features of the ordered state, like reduced anisotropy and pronounced magnetic non-linearity, are naturally interpreted by the presence of differently oriented magnetic domains.\\

\indent The work was partially performed at SINQ, Paul Scherrer Institute, Villigen, Switzerland. The sample preparation was supported by the NCCR research pool MaNEP of the Swiss NSF. The support of the Swiss NSF SCOPES project is gratefully acknowledged. We thank Dr. P.-J. Brown for fruitful discussions.
\end{document}